\documentclass{elsart}

\usepackage{graphicx}
\usepackage{amssymb}

\begin{document}

\begin{frontmatter}



\title{Blazar nuclei in radio-loud\\narrow-line Seyfert 1?}


\author[1]{L. Foschini},
\ead{foschini@iasfbo.inaf.it}
\author[2]{L. Maraschi},
\author[2]{F. Tavecchio},
\author[2]{G. Ghisellini},
\author[3]{M. Gliozzi} and
\author[4]{R. M. Sambruna}

\address[1]{INAF/IASF-Bologna, Via Gobetti 101, 40129 Bologna (Italy)}
\address[2]{INAF/Osservatorio Astronomico di Brera, Via Brera 28, 20121 Milano (Italy)}
\address[3]{George Mason University, 4400 University Drive, Fairfax, VA 22030 (USA)}
\address[4]{NASA/Goddard Space Flight Center, Code 661, Greenbelt, MD 20771 (USA)}

\begin{abstract} 
It has been suggested that some radio-loud narrow-line Seyfert 1 contain relativistic jets, on the basis of their flat-spectrum radio nuclei and studies on variability. We present preliminary results of an ongoing investigation of the
X-ray and multiwavelength properties of $5$ radio-loud NLS1 based on archival data from \emph{Swift} and \emph{XMM-Newton}. Some sources present interesting characteristics, very uncharacteristic for a radio-quiet narrow-line Seyfert 1, such as very hard X-ray spectra, and correlated optical and ultraviolet variability. However, none of the studied sources show conclusive evidence for relativistic jets. $\gamma-$ray observations with \emph{Fermi} are strongly recommended to definitely decide on the presence or not of relativistic jets.
\end{abstract}


\end{frontmatter}

\section{Introduction}
The radio-loudness phenomenon, i.e. the dominance of radio over optical emission, is still poorly understood. Basically, it is known that a small percentage of Active Galactic Nuclei (AGN) displays this type of activity, which is commmonly ascribed to the presence of a relativistic jet (Urry \& Padovani 1995), but the physics necessary to explain the generation, collimation, and evolution of the jet is yet to be known (e.g. Blandford 2001). Among the possible ways to investigate this problem, one is to study the population of radio-loud narrow-line Seyfert 1 galaxy (NLSy1). Indeed, NLSy1 -- as a peculiar subclass of Seyferts -- are generally radio-quiet, but a small percentage ($<7$\%, Komossa et al. 2006) of this population is radio-loud, \emph{i.e.}, with the ratio $R$ between the $5$~GHz and B filter ($4400$\AA) flux densities greater than $10$, according to the definition by Kellermann et al. (1989). Therefore, it is worth investigating this type of sources to understand the origin of the radio loudness and, if due to a relativistic jet, how this is possible in a population of active nuclei that are mostly radio quiet.

After a few isolated cases of radio-loud narrow-line Seyfert 1 galaxies (e.g. Padovani et al. 2002) and early surveys (e.g. Zhou \& Wang 2002, Komossa et al. 2006, Whalen et al. 2006), the advent of the \emph{Sloan Digital Sky Survey} (SDSS) made it possible to perform large scale surveys (Zhou et al. 2006, Yuan et al. 2008). Particularly, Komossa et al. (2006), with a sample including also sources with weak radio loudness ($R\gtrsim 10$), found that most of radio loud NLSy1 are resembling to GHz-Peaked/Compact Steep Spectrum sources (GPS/CSS), which in turn are believed to be the young stage of quasars (e.g. Guainazzi et al. 2006, Stawarz et al. 2008). Yuan et al. (2008) analyzed a sample of strong radio loud objects ($R\gtrsim 50$) and suggested similarities with high-energy peaked BL Lac Objects (HBL). Specifically, Yuan et al. (2008) claim to have found examples of ``high-frequency peaked flat spectrum radio quasars'' (HFSRQ, see Padovani 2007), which are quasars with synchrotron peak in the X-rays. This type of source is not foreseen by the classical ``blazar sequence'' (Fossati et al. 1998, Ghisellini et al. 1998) and therefore the discovery of such types of sources can represent a failure in the ``sequence''. Although some examples have been suggested (see Padovani 2007 for a review), no conclusive evidence has been presented. In addition, Ghisellini \& Tavecchio (2008) recently suggested that such type sources can exist and can be in agreement with the sequence by assuming that the jet is dissipating most of its energy beyond the broad-line region.

Therefore, since the study of radio-loud NLSy1 could have impact on the understanding of radio-loudness and the blazar sequence, we engaged in a study of a sample of these sources, by taking advantage of the power of \emph{XMM-Newton} and \emph{Swift} in giving simultaneous optical/UV/X-rays data. This allowed us to perform multiwavelength variability studies and to build spectral energy distributions (SED), which can be reproduced with some models generally applied to blazars. 

\begin{figure}
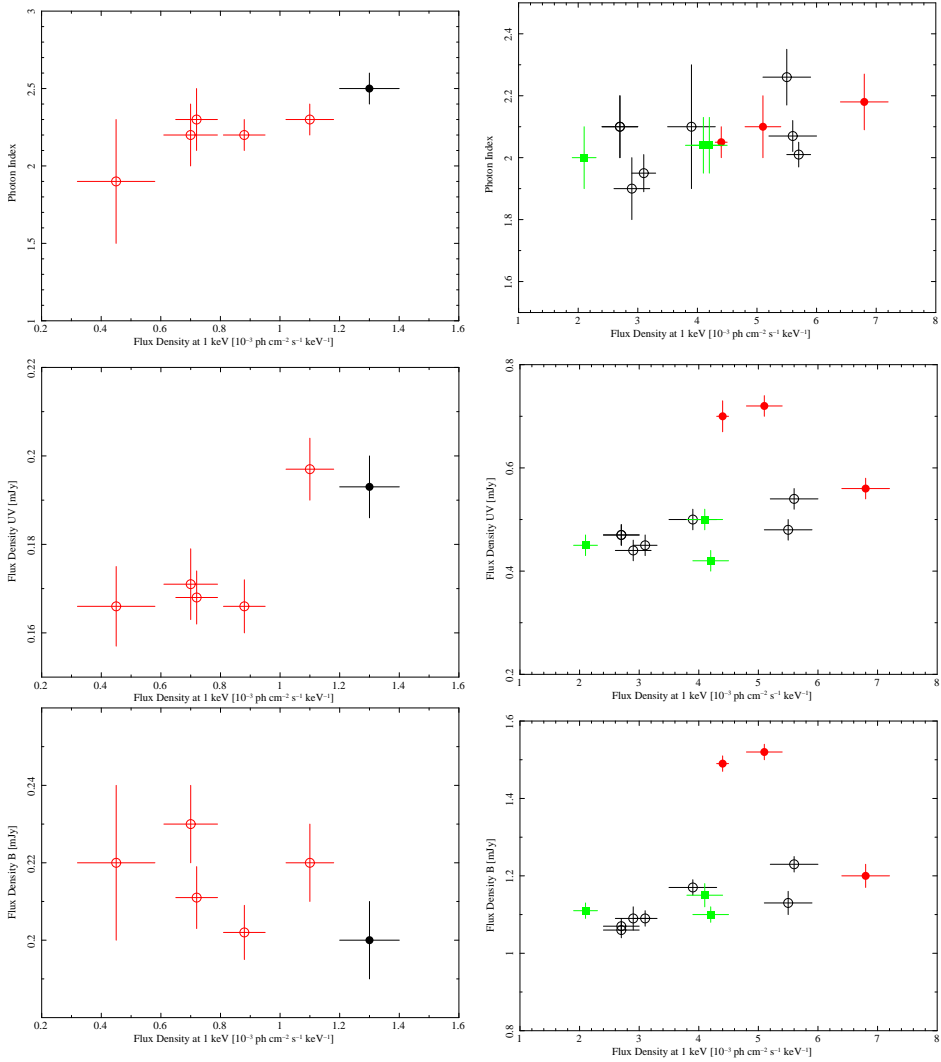

\begin{center}
\includegraphics*[angle=270,scale=0.26]{1629_gammaflux.ps}
\includegraphics*[angle=270,scale=0.26]{0323_gammaflux.ps}\\
\includegraphics*[angle=270,scale=0.26]{1629_fluxXfluxUV.ps}
\includegraphics*[angle=270,scale=0.26]{0323_fluxXfluxUV.ps}\\
\includegraphics*[angle=270,scale=0.26]{1629_fluxXfluxB.ps}
\includegraphics*[angle=270,scale=0.26]{0323_fluxXfluxB.ps}\\
\end{center}
\caption{\scriptsize $\Gamma$ (\emph{top panel}), UV (UVW1) flux density at $2634$~\AA (\emph{center panel}), and optical (B) flux density at $4329$~\AA (\emph{bottom panel}) versus X-ray flux density at $1$~keV for RGB~J$1629+401$ (\emph{left column}) and 1H~$0323+342$ (\emph{right column}). Filled circles indicate fits with a broken power-law model; open circles are instead for fits with a single power-law model; filled squares are for spectra with hints of features (emission lines).}
\label{fig:1629}
\end{figure}

\section{Sample selection and analysis}
We performed a cross-correlation between the radio-loud NLSy1 reported in Zhou et al. (2002) and Komossa et al. (2006), plus some specific source like 1H~$0323+342$ (Zhou et al. 2007), and the archives of publicly available data of \emph{XMM-Newton} and \emph{Swift}. We found data for 10 sources, but not all were useful. Indeed, the \emph{XMM-Newton} observation of IRAS~$01506+2554$ was severely affected by soft proton flares (anomalous high background) and no useful data remained after a proper screening. B3~$1702+457$ and MS~$1346.2+2637$ show a complex X-ray spectrum with emission lines and appear quite similar to typical Seyfert galaxies (indeed, their radio loudness is very low: $R=11$ for the former and $R=6-18$ for the latter). RX~J$2314.9+2243$ ($R=8-18$) has been observed twice by \emph{Swift}, but there is no useful statistics in one pointing and the other does not show any interesting feature. Last, but not least, PKS~$0558-504$ ($R=15-35$) has been observed $17$ times by \emph{XMM-Newton} and there are several papers already written on this source (Gliozzi et al. 2001, Brinkmann et al. 2004). In addition, this source will be the subject of a long multiwavelength campaign in the Autumn 2008 (PI Gliozzi). Therefore, it is not analyzed here. 

Five sources remain: RGB~J$1629+401$ ($R=35-182$), 1H~$0323+342$ ($R=38-151$), RX~J$0134.2+4258$ ($R=36-178$), SDSS~J$172206.03+56541.6$ ($R=70-773$), PKS~$2004-447$ ($R=1710-6320$), for which we found \emph{Swift} data, except for the last one (\emph{XMM-Newton}). The analysis was performed by using standard methods (e.g. Foschini et al. 2008 for \emph{XMM-Newton} and Foschini et al. 2007 for \emph{Swift}), but with the latest version of the software and the calibration data base (\texttt{HEASoft 6.5} and \texttt{SAS 8.0}). 

\begin{figure}
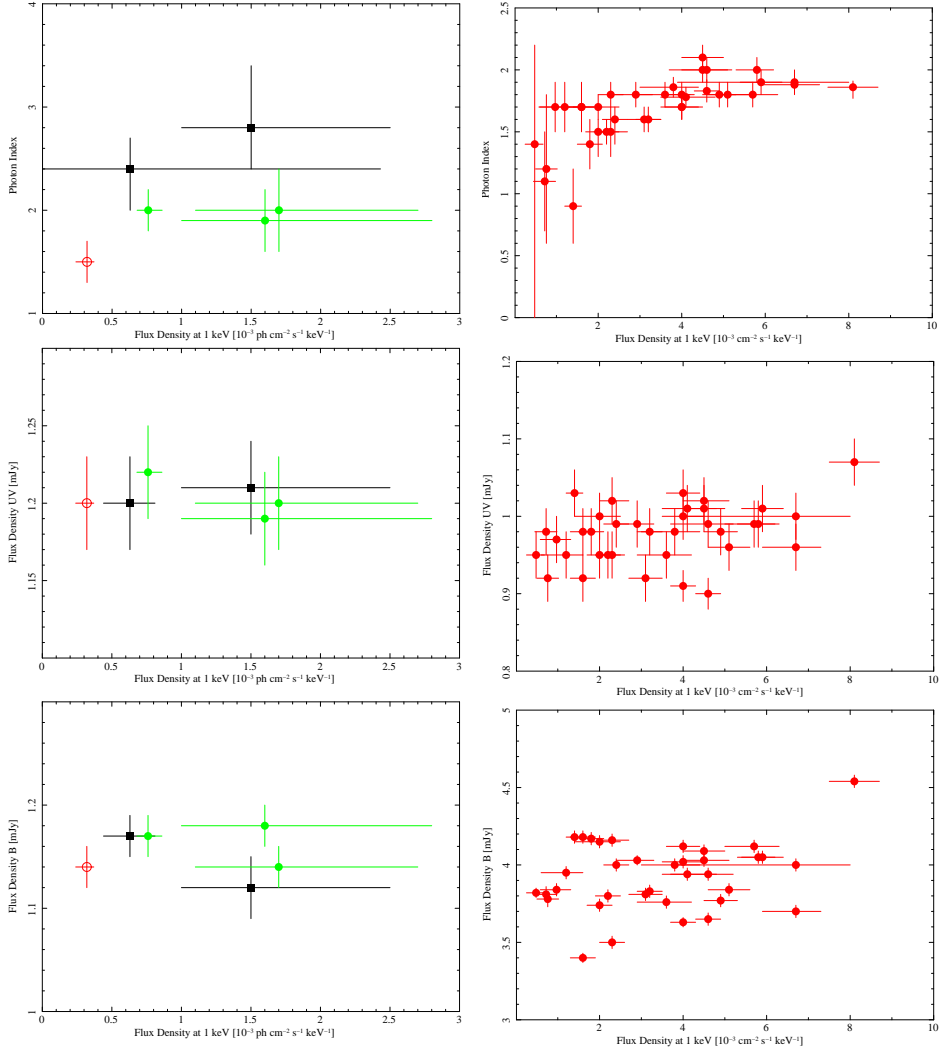

\begin{center}
\includegraphics*[angle=270,scale=0.26]{0134_gammaflux.ps}
\includegraphics*[angle=270,scale=0.26]{mkn766_gammaflux.ps}\\
\includegraphics*[angle=270,scale=0.26]{0134_fluxXfluxUV.ps}
\includegraphics*[angle=270,scale=0.26]{mkn766_fluxXfluxUV.ps}\\
\includegraphics*[angle=270,scale=0.26]{0134_fluxXfluxB.ps}
\includegraphics*[angle=270,scale=0.26]{mkn766_fluxXfluxB.ps}\\
\end{center}
\caption{\scriptsize $\Gamma$ (\emph{top panel}), UV (UVW1) flux density at $2634$~\AA (\emph{center panel}), and optical (B) flux density at $4329$~\AA (\emph{bottom panel}) versus X-ray flux density at $1$~keV for RX~J$0134.2-4258$ (\emph{left column}) and Mkn~$766$ (\emph{right column}), which is presented as a template of narrow-line Seyfert 1 radio quiet. Filled circles indicate fits with a broken power-law model; open circles are instead for fits with a single power-law model; filled squares are for spectra with hints of a warm absorber.}
\label{fig:MKN766}
\end{figure}

\begin{figure}
\begin{center}
\includegraphics*[angle=270,scale=0.4]{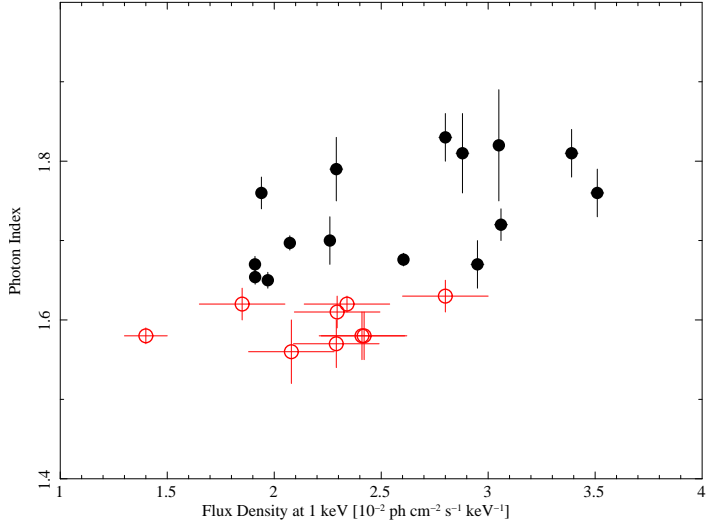}
\end{center}
\caption{\scriptsize X-ray flux density at $1$~keV versus $\Gamma$ for 3C~$273$. Black filled circles refer to \emph{XMM-Newton} data from Foschini et al. (2006a) and cover the years $2000-2004$; while open circles refer to \emph{BeppoSAX} data from Grandi \& Palumbo (2004), covering the years $1996-2001$. During the \emph{BeppoSAX} years, there was high radio flux, the source was jet dominated and the contribution from the Seyfert component was low. The contrary occurred during the \emph{XMM-Newton} period.}
\label{fig:3C273}
\end{figure}

\section{Discussion}
Fig.~\ref{fig:1629} and \ref{fig:MKN766}, (\emph{left column}) show the X-ray flux density at 1 keV versus $\Gamma$, UV and optical flux densities of RGB~J$1629+401$, 1H~$0323+342$, RX~J$0134.2+4258$. To compare with the behavior of a typical NLSy1 radio-quiet, we have analyzed the data of Mkn~$766$. We note that, as expected, Seyfert X-ray spectra become steeper with higher fluxes (see Fig.~\ref{fig:MKN766}, \emph{right column}; cf with, e.g., Papadakis et al. 2002). Blazars show harder spectra with increasing flux, but this is not true for all the types of blazars. For example, Fig.~\ref{fig:3C273} shows the X-ray flux density at $1$~keV versus $\Gamma$ for 3C~$273$, a source thought to contain both a Seyfert-like nucleus and a relativistic jet (Grandi \& Palumbo 2004). In this case, the X-ray spectrum displays a steepening with increasing flux, although the photon index is generally harder than that typical of a NLSy1. Therefore, the evolution of the X-ray spectrum alone is not sufficient to distinguish between a blazar-like or Seyfert-like behaviour. 

It is necessary to analyze also the radiation at optical/UV wavelengths. In blazars, the emission at these frequencies is dominated by the jet and is therefore correlated with X-rays. In Seyferts, the optical/UV radiation is dominated by the accretion disk and the host galaxy, while the X-ray continuum is generated by a thermally Comptonized corona. Therefore, different pattern of variability and correlations are expected (e.g. Nandra \& Papadakis 2001). 

Among the sources analyzed in the present work, 1H~$0323+342$ is the most interesting case: it shows some correlation with optical and UV fluxes at the highest X-ray fluxes, where the X-ray spectra can be well fitted with a broken power-law model extending to hard X-rays (Fig.~\ref{fig:1629}, \emph{right column}; in these cases, indicated with a filled circle, the figure displays only the soft photon index). Indeed, this source has been already detected by \emph{INTEGRAL} (Bird et al. 2007, Krivonos et al. 2007, Malizia et al. 2007), but it was indicated as a normal Seyfert active nucleus and none underlined the peculiar properties of this source. We reanalyzed with \texttt{OSA 7.0} all the publicly available \emph{INTEGRAL}/IBIS data, for a total exposure of $\approx 200$~ks (performed in $2004$, revolution $220$), and we detected the source at $\approx 2.5$~mCrab flux level in the $20-40$~keV energy band ($5\sigma$ significance), but there was no detection in the $40-100$~keV band, with an upper limit of $2.6$~mCrab. We also detected the source in the \emph{Swift}/BAT data (exposure $\approx 53$~ks, performed in $2006-2008$), with a flux of $\approx 16$~mCrab in the $40-100$~keV band ($8\sigma$ significance) and no detection in the $20-40$~keV band, with an upper limit of $20$~mCrab. Therefore, 1H~$0323+342$ shows a strong flux and spectral variability: low flux and soft spectrum during the \emph{INTEGRAL} observation and high flux and hard spectrum during the \emph{Swift} pointings. 

\begin{figure}
\begin{center}
\includegraphics*[scale=0.6,clip,trim=0 50 0 60]{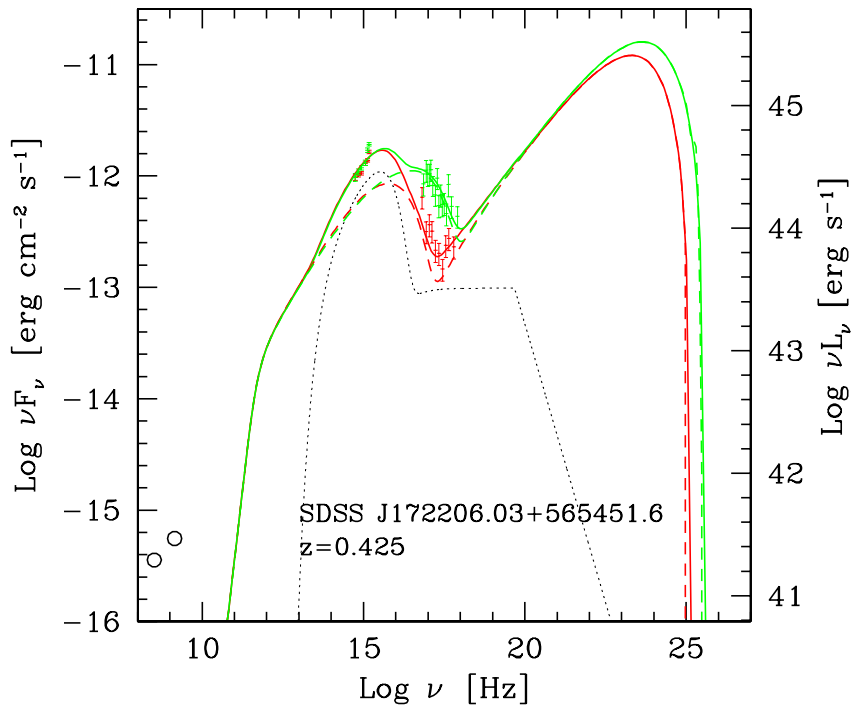}
\end{center}
\caption{\scriptsize Spectral energy distribution (SED) of SDSS~J$172206.03+56541.6$, with a possible model (continuous lines) including disk (dotted line) and jet emission (dashed lines). Green and red colors indicate two different states of the source.}
\label{fig:SDSS}
\end{figure}

The presence of such a hard X-ray component, together with an increase of optical/UV fluxes, can be likely attributed to a jet. 1H~$0323+342$ appears to be like a X-ray selected flat-spectrum radio quasar (FSRQ, see Maraschi et al. 2008), where, during quiescence, the X-ray emission is dominated by a corona and the jet contribution is negligible, while the contrary occurs during outbursts. The jet radiation overwhelming the corona is indicated by the emergence of the hard power-law component. It is not possible to make a firm statement, given the low statistics, but surely this behaviour is very uncharacteristic for a narrow-line Seyfert 1.

Also RX~J$0134.2+4258$ shows a hard X-ray spectrum ($\Gamma = 1.5\pm 0.2$) in one pointing, while the remaining pointings display the typical properties of a radio-quiet narrow-line Seyfert 1 (Fig.~\ref{fig:MKN766}, \emph{left column}). 

The case of PKS~$2004-447$ has been extensively studied by Gallo et al. (2006), who also noted a blazar-like behavior of the source, possibly with an additional Seyfert component. The X-ray spectrum is well fitted with a broken power-law model with $\Gamma_{soft}=2.0\pm 0.2$, $\Gamma_{hard}=1.49\pm 0.03$ and $E_{break}=0.66\pm 0.08$~keV. We note that the soft component ($0.2-1$~keV) is strongly variable ($RMS=16\pm 4$\%), while the hard one ($2-10$~keV) is consistent with a constant ($RMS < 8$\%, $3\sigma$ upper limit). This suggests a different origin of the two components and a behavior
similar to Low-Frequency peaked BL Lacs (LBL, cf, \emph{e.g.}, Foschini et al. 2006b). 

A similar case is SDSS~J$172206.03+56541.6$. There are two \emph{Swift} observations with significant changes in the X-ray spectrum. On June $23$, $2007$ data, the best fit is obtained using a broken power-law model ($\Gamma_{soft}=2.8_{-0.4}^{+0.7}$, $\Gamma_{hard}=1.5_{-0.6}^{+0.5}$, $E_{break}=1.0\pm 0.4$~keV), while $11$ days later the X-ray spectrum is best fitted with a single power-law model ($\Gamma=2.4\pm 0.1$). The X-ray and UV fluxes increase from the first observation to the second one. We decided to try a fit of the SED built with these two \emph{Swift} observations (Fig.~\ref{fig:SDSS}) and we adopted a modified synchrotron self-Compton (SSC) model by Maraschi \& Tavecchio (2003) to which we refer the reader for more details. In the present model, the optical/UV/X-ray emission is generated by a disk plus a jet. The distribution of electrons is shaped with a broken power-law with index $n_1$ from $\gamma_{min}$ to $\gamma_{break}$ and index $n_2$ from the break up to $\gamma_{max}$.  The parameters resulting from the fit on the two data sets of SDSS~J$172206.03+56541.6$ are basically the same, except for the $\gamma_{max}$, which changes from $4\times 10^4$ to $10^5$, and $n_2$, which decreases from $3.6$ to $3.2$. 

The X-ray emission of the low state (the broken power-law model indicated above) is modeled with a synchrotron plus inverse-Compton components: the former takes into account the soft X-ray emission up to $E_{break} \sim 1$~keV, while the latter can explain the hard X-ray radiation above that energy. In the active state, the peak of the synchrotron emission moves to higher frequencies and the X-ray energy band is dominated by this process, the behaviour expected from a HBL. Another possibility is that the jet is dissipating its energy beyond the narrow-line region in NLSy1, like in the case of RGB~J$1629+401$ (see below). Obviously, the final word on this modeling can be written if and when \emph{Fermi} will detect it: according to our estimates, the source could be already detected after a few months of exposure. 

In the case of RGB~J$1629+401$ there are some hints of correlation of the highest X-ray fluxes with high UV flux (Fig.~\ref{fig:1629}, \emph{left column}). For the modeling, it was assumed that the dissipation of the jet occurs outside the broad-line region and therefore the non-thermal emission can be modeled with a synchrotron self-Compton (SSC) model. See Maraschi et al. (2008) for more details.

\section{Conclusions}
The analysis of the optical-to-X-ray data presented here suggests that in the class of radio-loud narrow-line Seyfert 1 galaxies there are some sources with interesting behavior, in some way similar to blazars. There is no ``unified way'' for radio-loud NLSy1 to be similar to blazars: basically, in most of the analyzed cases, the X-ray spectrum, generally soft, shows a break with the emergence of a hard component. In one case, the X-ray spectrum can be fitted with a single power-law model with a very hard photon index. Therefore, it appears that there is a Seyfert component (represented by a disk plus a X-ray corona), but sometimes the X-ray spectrum shows an evident change of the shape, together with correlated changes in optical and ultraviolet fluxes, suggesting the strengthening of a jet emission. This occurred at least twice for 1H~$0323+342$ and at least once for RX~J$0134.2+4258$, PKS~$2004-447$. The observed changes in spectral shapes and fluxes can be interpreted in the framework of the aborted jets (Ghisellini et al. 2004), but considering that sometimes the jet is launched successfully and then the source becomes radio loud. 

Another possibility is that the jet dissipates most of its energy beyond the narrow-line region, like the blue quasars of Ghisellini \& Tavecchio (2008). The cases of RGB~J$1629+401$ and SDSS~J$172206.03+56541.6$ can be explained by this theory.

The flat radio spectra reported by other authors (Zhou et al. 2002, Komossa et al. 2006) support the interpretation of a jet origin of part of the emitted radiation, but the data available are not conclusive. $\gamma-$ray detection of these sources with \emph{Fermi} above $100$~MeV would be the decisive signature of the presence of a relativistic jet.

\textbf{References}\\
\scriptsize
Bird, A. J., Malizia, A., Bazzano, A., et al., The Third IBIS/ISGRI Soft Gamma-Ray Survey Catalog, ApJS, 170, 175-186, 2007.\\
Blandford, R. D., Particles and fields in radio galaxies: a summary, ASP Conf.~Ser.~250: Particles and Fields in Radio Galaxies Conference, 250, eds. Laing, R.A \& Blundell, K.M., p. 487, 2001\\
Brinkmann, W., Ar\'evalo, P., Gliozzi, M., Ferrero, E., X-ray variability of the narrow line Seyfert 1 galaxy PKS~$0558-504$, A\&A, 415, 959-969, 2004.\\
Foschini, L., Ghisellini, G., Raiteri, C.M., et al., XMM-Newton observations of a sample of $\gamma-$ray loud active galactic nuclei, A\&A, 453, 829-838, 2006a.\\
Foschini, L., Tagliaferri, G., Pian, E., et al., Simultaneous X-ray and optical observations of S5~$0716+714$ after the outburst of March~$2004$, A\&A, 455, 871-877, 2006b.\\
Foschini, L., Ghisellini, G., Tavecchio, F., et al., X-ray/UV/Optical follow-up of the blazar PKS~$2155-304$ after the giant TeV flares of $2006$~July, ApJ, 657, L81-L84, 2007.\\
Foschini, L., Treves, A., Tavecchio, F., et al., Infrared to X-ray observations of PKS~$2155-304$ in a low state, A\&A, 484, L35-L38, 2008.\\
Fossati, G., Maraschi, L., Celotti, A., Comastri, A., \& Ghisellini, G., A unifying view of the spectral energy distribution of blazars, MNRAS, 299, 433-448, 1998.\\
Gallo, L., Edwards, P.G., Ferrero, E., et al., The spectral energy distribution of PKS~$2004-447$: a compact steep-spectrum source and possible radio-loud narrow-line Seyfert 1 galaxy, MNRAS, 370, 245-254, 2006.\\
Ghisellini, G., Celotti, A., Fossati, G., Maraschi, L., \& Comastri, A., A theoretical unifying scheme for gamma-ray bright blazars, MNRAS, 301, 451-468, 1998.\\
Ghisellini, G., Haardt, F., Matt, G., Aborted jets and the X-ray emission of radio-quiet AGNs, A\&A, 413, 535-545, 2004.\\
Ghisellini, G., \& Tavecchio, F., The blazar sequence: a new perspective, MNRAS, 387, 1669-1680, 2008.\\
Gliozzi, M., Brinkmann, W., O'Brien, P.T., et al., XMM-Newton monitoring of X-ray variability in the quasar PKS~$0558-504$, A\&A, 365, L128-L133, 2001.\\
Grandi, P., \& Palumbo, G.G.C., Jet and accretion-disk emission untangled in 3C 273, Science, 306, 998-1002, 2004.\\
Guainazzi, M., Siemiginowska, A., Stanghellini, C., et al., A hard X-ray view of giga-hertz peaked spectrum radio galaxies, A\&A, 446, 87-96, 2006.\\
Kellermann, K. I., Sramek, R., Schmidt, M., Shaffer, D. B., Green, R., VLA observations of objects in the Palomar Bright Quasar Survey, AJ, 98, 1195-1207, 1989.\\
Komossa, S., Voges, W., Xu, D., et al., Radio-loud narrow-line type 1 quasars, AJ, 132, 531-545, 2006.\\
Krivonos, R., Revnivstev, M., Lutovinov, A., Sazonov, S., Churazov, E., Sunyaev, R., INTEGRAL/IBIS all-sky survey in hard X-rays, A\&A, 475, 775-784, 2007.\\
Malizia, A., Landi, R., Bassani, L., et al., Swift/XRT Observation of 34 New INTEGRAL/IBIS AGNs: Discovery of Compton-Thick and Other Peculiar Sources, ApJ, 668, 81-86, 2007.\\
Maraschi, L., \& Tavecchio, F., The jet-disk connection and blazar unification, ApJ, 593, 667-675, 2003.\\
Maraschi, L., Foschini, L., Ghisellini, G., Tavecchio, F., Sambruna, R.M., Testing the blazar spectral sequence: X-ray selected blazars, MNRAS, accepted for publication, 2008, [\texttt{arXiv:0810.0145}].\\
Nandra, K. \& Papadakis, I.E., Temporal Characteristics of the X-Ray Emission of NGC 7469, ApJ, 554, 710-724, (2001).\\
Padovani, P., Costamante, L., Ghisellini, G., Giommi, P., Perlman, E., BeppoSAX observations of synchrotron X-ray emission from radio quasars, ApJ, 581, 895-911, 2002.\\
Padovani, P., The blazar sequence: validity and predictions, Ap\&SS, 309, 63-71, 2007.\\
Papadakis, I.E., Petrucci, P.O., Maraschi, L., et al., Long-Term Spectral Variability of Seyfert Galaxies from Rossi X-Ray Timing Explorer Color-Flux Diagrams, ApJ, 573, 92-104, 2002.\\
Stawarz, {\L}., Ostorero, L., Begelman, M.C., et al., Evolution of and High-Energy Emission from GHz-peaked spectrum sources, ApJ, 680, 911-925, 2008.\\
Urry, C.M., \& Padovani, P., Unified Schemes for Radio-Loud Active Galactic Nuclei, PASP, 107, 803-845, 1995.\\
Whalen, D.J., Laurent-Muehleisen, S.A., Moran, E.C., Becker, R.H., Optical properties of radio-selected narrow-line Seyfert 1 galaxies, AJ, 131, 1948-1960, 2006.\\
Yuan, W., Zhou, H.Y., Komossa, S., et al., A population of radio-loud narrow-line Seyfert 1 galaxies with blazar-like properties?, ApJ, 685, 801-827, 2008.\\
Zhou, H.-Y., \& Wang, T.-G., Properties of broad band continuum of narrow line Seyfert 1 galaxies, Ch. J. A\&A, 2, 501-524, 2002.\\
Zhou, H.-Y., Wang, T.-G., Yuan, W., et al., A comprehensive study of 2000 narrow-line Seyfert 1 galaxies from the Sloan Digital Sky Survey. I. The sample, ApJSS, 166, 128-153, 2006.\\
Zhou, H.-Y., Wang, T.-G., Yuan, W., et al., A narrow-line Seyfert 1 - Blazar composite nucleus in 2MASX J0324+3410, ApJ, 658, L13-L16, 2007.\\

\end{document}